\documentclass[useAMS,usenatbib]{mn2e}
\usepackage{graphicx}
\usepackage{times}
\title[Emission line activity in type 2 quasars]{Emission line activity in type 2 quasars from the Sloan Digital Sky Survey}
\author[Villar-Mart\'\i n et al.]{M. Villar-Mart\'\i n$^1$, A. Humphrey$^{2,3}$, 
A. Mart\'\i nez-Sansigre$^4$, M. P\'erez-Torres$^1$
\newauthor   L. Binette$^{3,5}$, X.G. Zhang$^{3,6}$ \\
$^{1}$Instituto de Astrof\'\i sica de Andaluc\'\i a (CSIC), Aptdo. 3004,
Granada, Spain\\
$^{2}$Korea Astronomy and Space Science Institute, 61-1, Hwaam-dong, Yuseong-gu, Daejeon, Republic of Korea 305-348\\
$^{3}$Instituto de Astronom\'\i a, Universidad Nacional Aut\'onoma de M\'exico, Ap. 70-264, 04510, DF, M\'exico\\
$^{4}$Max-Planck-Institut f\"ur Astronomie, K\"onigstuhl 17, D-69117 Heidelberg, Germany\\
$^{5}$D\'epartement de Physique, de G\'enie Physique et d'Optique, Universit\'e Laval, Qu\'ebec, QC, G1K 7P4, Canada\\
$^{6}$Max-Planck-Institut f\"ur Astrophysik, Karl Schwarschild-Str. 1, 85748 Garching, Germany}

\begin{document}

\date{Accepted 2008 July 29.  Received 2008 July 29; in original form 2008 February 1}

\pagerange{\pageref{firstpage}--\pageref{lastpage}} \pubyear{2002}

\maketitle

\begin{abstract}
We have compared the optical emission line ratios of type 2 quasars from
Zakamska et al.
with standard AGN photoionization model predictions, Seyfert 2s, HII galaxies,
and narrow line FRII radio galaxies.   Moderate to high ionization narrow line radio galaxies and Seyfert 2s are indistinguishable from type 2 quasars based 
on their optical line ratios. 
The standard AGN photoionization models, widely discussed for other type 2 AGNs, can reproduce successfully the loci and trends
of type 2 quasars in some of the  main diagnostic diagrams.
These models are not exempt of problems and the discrepancies with the data
are the same encountered for other type 2 AGNs. As for these, realistic 
 models must take into account a range of cloud properties, as widely demonstrated in the literature.

The Zakamska et al. sample is strongly biased towards objects with  high line luminosities (L[OIII]$>$10$^{42}$ erg s$^{-1}$). 
We have found that stellar photoionization is obvious  in a  fraction of objects (3 out of 50)
which are characterized by low [OIII] luminosities compared with most type 2 quasars in the sample. We suggest that if the sample were expanded towards
lower line luminosities ($\sim$10$^{40-42}$ erg s$^{-1}$) stellar photoionization might be evident  much more frequently.

We explore an alternative scenario to pure AGN photoionization
in which a varying contribution of stellar ionized gas contributes to
the line fluxes. Although the models presented here are rather simplistic and not strong quantitative results
can be extracted regarding the relative contribution of stellar vs. AGN photoionization, 
our results suggest that adding a varying contribution of stellar photoionized gas works in the right direction 
to solve most of the problems affecting the standard AGN photoionization models.
The ``temperature problem'', on the other hand remains.

 \end{abstract}


\label{firstpage}

\section{Introduction}

Type 2 active galactic nuclei (AGNs) are those AGNs whose permitted and
forbidden lines have similar values of full width at half maximum (FWHM). In many cases (e.g. Seyfert 2 galaxies) this can be explained as a consequence of obscuration of the central region by large column densities of gas and dust. In the standard unification model the obscuring
structure is toroidal in shape so that the view to the inner nuclear region
is blocked for some orientations (Antonucci \citeyear{ant93}). According to this model,
certain classes of type 1 and type 2 AGNs are the same entities, but have different orientations relative to the 
observer line of sight.
If this model is valid for the most luminous AGNs (quasars), there
must exist a high-luminosity family of type 2 quasars. 

The existence of such an
object class was predicted a long time ago, but it has been only in the last few years that type 2 quasars  have been discovered
in large quantities in X-ray, mid-IR  (Mart\'\i nez-Sansigre  et al. \citeyear{mar05}, Szokoly  et al. \citeyear{szo04}) and
optical surveys (Zakamska et al.  \citeyear{zak03}).   \cite{zak03}
 identified  $\sim$145 objects in the redshift range
0.3$\la z \la$0.8 in the Sloan Digital Sky Survey (SDSS; York et al. \citeyear{york00})  with the high ionization
narrow emission line spectra characteristic of type 2 AGNs and  narrow
line luminosities typical of type 1 quasars (see also Reyes et al. \citeyear{rey08}). 
 Their IR and X-ray properties are consistent
with their interpretation as powerful obscured AGN (Ptak et al. \citeyear{ptak06},  Zakamska et al. \citeyear{zak04}). They  show a wide range of X-ray luminosities and obscuring column densities.  About 40 objects in their sample were detected with IRAS and
have infrared luminosities among the most luminous quasars at similar redshift.
The host galaxies are ellipticals, although with irregular morphologies, and
the nuclear optical emission is highly polarized (Zakamska  et al. \citeyear{zak06}). The detection rate
in radio ($\sim$10\%, Zakamska et al. \citeyear{zak04}; Vir Lal  \& Ho \citeyear{vir07}) is consistent with that of other AGN types.

Spectroscopic studies of type 2 quasars have focused
so far on identifying   emission lines, measuring
 some basic parameters 
(line luminosities, redshift, line widths)  and searching for correlations
among them and with other observables (equivalent widths,
color magnitudes, radio luminosities, etc). All this is critical to classify
 the objects, to investigate the nature of the powering
mechanism, the  obscuring structure and, ultimately, test the validity
of the unification models (e.g. Reyes et al. \citeyear{rey08}). 

However,  little work has been done
to characterize the gaseous and ionization properties 
of type 2 quasars. Given this  lack of knowledge, 
the primary goal of the work presented here is to use the emission line information to study  the excitation mechanism, the physical conditions and ionization properties of the gas. 

 Detailed spectroscopic studies of other active galaxy types, such as narrow line radio galaxies (NLRGs, e.g. Robinson et al. \citeyear{rob87}) have allowed during the last few decades  the characterization of the
chemical abundances of the gas, the ionization mechanism, 
the physical and kinematic properties, etc. Similar work must be done for type 2 quasars.  These studies will ultimately provide valuable information
about the formation process of the host galaxy, the star forming and chemical
enrichment  histories and the origin of the nuclear activity (e.g. Tadhunter, Fosbury \& Quinn \citeyear{tadh89}, Villar-Mart\'\i n et al. \citeyear{vm05}, Humphrey et al. \citeyear{hum08}).
 An important disadvantage of  NLRG  over type 2 quasar studies
 is that  the radio  activity, via jet-gas interactions, imprints  important distortions on the  properties of the ionized gas making it difficult to investigate  the intrinsic properties of the host galaxy and environment, as well as the chemical abundances and
the nature of the excitation mechanism (e.g. Tadhunter \citeyear{tadh02}). The study of radio-quiet type 2 quasars should not suffer from such effects.

Throughout this paper we assume  $\Omega_{\Lambda} =$ 0.73, $\Omega_{m}$ = 0.27 and $H_{0}$ = 71 km  s$^{-1}$  Mpc$^{-1}$.

\section{The object sample}

The objects studied in this paper are a sub-sample of candidate
type 2 quasars from 
the Sloan Digital
Sky Survey (SDSS) selected by \cite{zak03}
in the redshift range 0.3$\la z \la$0.8 ($\sim$145 objects). The selection criteria 
applied by  \cite{zak03} are listed below (notice that not necessarily all criteria apply to all objects. See Zakamska et al. \citeyear{zak03} for more 
detailed information)).\footnote{Throughout the paper the emission lines will be named as follows:
[OIII] for [OIII]$\lambda$5007; [OII] for [OII]$\lambda$3727;
[NII] for [NII]$\lambda$6583; [NeV] for [NeV]$\lambda$3426; [NeIII] for [NeIII]$\lambda$3689; HeII for HeII$\lambda$4686.}

\begin{itemize}

\item $i\le$19.1

\item 0.3 $<z<$ 0.83

\item Signal to noise ratio $>$7.5

\item Equivalent width (EW) of [OIII]$\ge$4 \AA

\item Luminosity of [OIII], L[OIII]$\ge$3$\times$10$^8$ L$_ {\odot}$

\item Full width at half maximum, FWHM(H$\beta$)$<$2000 km s$^{-1}$

\item Selection of active galaxies among  emission line galaxies, 
based on line ratio criteria (different
line ratios depending on $z$)

\item  Other signs of AGN activity such as the detection of the high ionization lines  [NeV]$\lambda\lambda$346,3426 and/or FWHM([OIII])$>$400 km s$^{-1}$.

\end{itemize}

  The 
sub-sample studied here  contains  50 type 2 quasars, which
 have been selected 
to span the full $z$ range  and the full [OIII] luminosity
range ($\ge$3$\times$10$^8$ L$_ {\odot}$) of the original type 2 quasar sample.  We did not set constraints  on the line
equivalent widths, although the selection criteria on the original sample do
contain an EW criterion, as stated above.

The  spectra were corrected for Galactic extinction.
 A galaxy template spectrum was subtracted for objects
with low line EWs, to correct for possible underlying stellar
and interstellar absorption (Zhang, Dultzin-Hacyan \& Wang \citeyear{zh07}).
 This procedure was necessary for a  small fraction
of objects ($\sim$15\% only).

 The  spectra were not corrected for internal dust reddening because 
such correction was  not possible for all objects. It should not affect our conclusions
since we will compare our results with previous
works on other AGN types (Seyfert 2s, radio galaxies), in which no internal extinction correction
was applied either.

\section{Analysis and results}

We investigate in this section the dominant ionizing mechanism of the optical line emitting
gas in the type 2 quasar sub-sample: AGN vs. stellar photoionization.

We will ignore shocks in our discussion as an alternative ionization mechanism
(e.g. Dopita \& Sutherland \citeyear{dop96}). In our sub-sample,  44 out of the 47 objects for which  radio information is
available are radio quiet  (i.e. $L_{1.4 GHz}<$10$^{31}$ erg s$^{-1}$ Hz$^{-1}$ sr$^{-1}$) and therefore, shocks induced by the radio structures are
not expected to play a significant role in the ionization of the gas (e.g. Clark et al. \citeyear{clark98}, Villar-Mart\'\i n et al. \citeyear{vm99}).
In the vast majority of the type 2 quasars in the sample considered
here, the
 bulk of the line profiles is characterized by rather quiescent kinematics, rather than perturbed, as one would expect if shocks were present.

\subsection{AGN photoionization}

We show in Fig.~1 several diagnostic diagrams involving optical emission lines
in which we plot the location of the SDSS type 2 quasar sub-sample (green, solid circles and blue solid triangles).
For comparison, we plot in the same diagrams the locus of HII galaxies
 from the catalogue of  \cite{ter91}  (magenta small symbols). These are
 absent in diagrams involving the HeII and [NeV]  because such lines
are rarely detected in this object class.

\begin{figure*}
\includegraphics{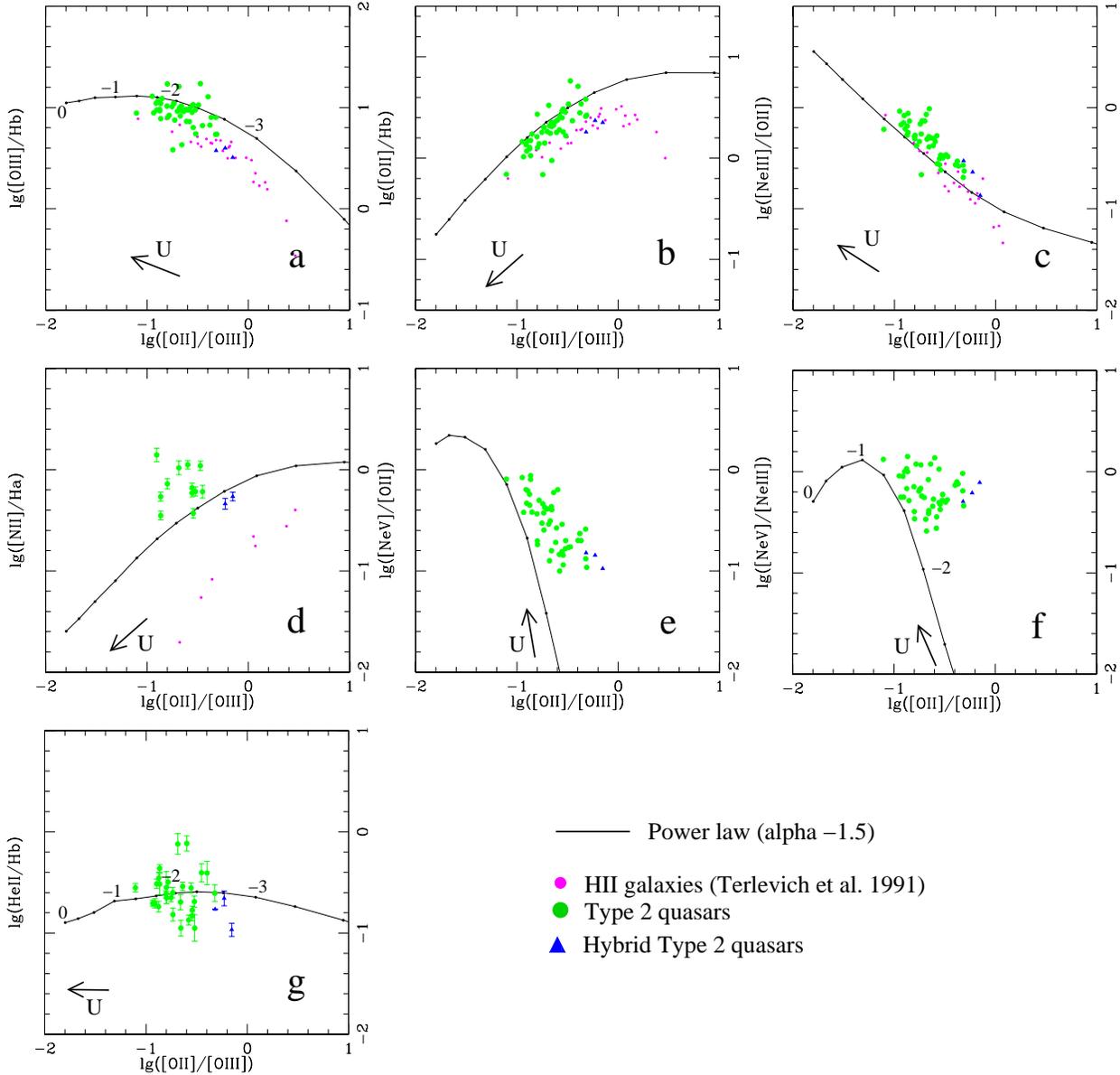}
\vspace{6.5in}
\caption{Loci of type 2 quasars    in
diagnostic diagrams involving the main optical emission lines detected in most spectra.  Hybrid objects are those with signatures of stellar
photoionization.
The solid line is the  standard sequence of solar metallicity
AGN photoionization models along which the ionization parameter $U$ varies. 
 lg($U$) values are shown for some models in several diagrams. The arrows indicate the sense of increasing $U$.  Errorbars are shown in those diagrams where the
measurement errors could have an impact on our interpretation and conclusions ($d$ and $g$). In all other diagrams the errorbars are in general smaller.
The comparison with the models and HII galaxies suggests that in general, AGN photoionization plays an important
role in ionizing the gas and the ionization parameter varies from object to object. On the other hand,
the standard AGN sequence shows some discrepancies with the data (e.g. diagrams $d$, $e$, $f$, $g$) which are the same encountered for other
type 2 AGNs.
 Stellar photoionization is apparent in
a small fraction of objects (hybrid object).}
\end{figure*}

The  solid lines  represent the standard \footnote{$U=\frac{Q}{4~\pi~r^2~n_H~c}$, where $Q$ is the ionizing photon luminosity of the source, $r$ is the distance between the cloud and the ionizing source, $n_H$ is the hydrogen density at the illuminated face of the cloud and $c$ is the speed of light.}$U$ sequence of photoionization models, built with the multipurpose code MAPPINGS Ic (Binette, Dopita \& Tuohy \citeyear{bin85}; Ferruit et al. \citeyear{fer97})
that  reproduces some of the main properties of the emission line spectra of  narrow line
 radio galaxies at different redshifts
(e.g. Robinson et al. \citeyear{rob87}, Humphrey et al. \citeyear{hum08}). The ionizing continuum
is a power law of index $\alpha$=1.5 ($F_{\nu} \propto \nu^{-\alpha}$), with a cut off energy  of 50 keV. The clouds  are considered to be  isobaric, plane-parallel, dust-free
ionization-bounded  slabs of density $n=$100 cm$^{-3}$  at the
illuminated face and characterized by solar abundances (Anders \& Grevesse \citeyear{anders89}). Since the SDSS spectra integrate
the emission from  nuclear (i.e, the narrow line region, NLR) and (possibly)  extended gas, a range of cloud densities
is expected with values as high as 10$^{6}$ cm$^{-3}$ or more. 
 Given the low
critical density of [OII]$\lambda$3727 ($\sim$3000 cm$^{-3}$), such high
densities will pose difficulties  to reproduce the observed strength
of [OII] relative to other emission lines, which suggest a  substantial 
contribution to the integrated spectrum of a low density component. Thus,
although it is clear that assuming a single density is a rough simplification (characteristic
on the other hand of the standard AGN sequence discussed in numerous papers), we will
assume $n_e$=100 cm$^{-3}$ as a representative density of the gas within the SDSS aperture
 and mention the possible impact of higher
densities when necessary.

There are several diagrams which are specially sensitive to the ionization
parameter $U$, in the sense that for a specific ionization mechanism, 
the line ratios involved are strongly influenced by a change 
in $U$. These   diagrams are $a$, $b$, $c$, $e$ and $f$ (Fig.~1). The  
 models shows the large shift along the sequence expected
as $U$ varies.
Looking at the distribution of data points in diagrams $a$, $b$, $c$ and $e$,  
type 2 quasars follow 
 a sequence  whose shape  can be accounted for by a variation in $U$. Most objects
have lg($U$) values in the range [-2.7,-1.7]

 There are several arguments that suggest that AGN photoionization plays an important role
in ionizing the gas:

\begin{itemize}

\item The shape and location of the sequence defined by type 2 quasars is adequately accounted for by the AGN models in the $a$, $b$ and $c$ diagrams.

\item Type 2 quasars have similar line ratios than Seyfert 2 and narrow line radio galaxies.
We compare in Fig.~2 the loci of type 2 quasars, 
Seyfert 2 (black open triangles) and radio galaxies (red diamonds for the
extended emission line regions (EELR) and red 'x' for the nuclear regions) 
 in the   [OIII]/H$\beta$ vs. [OII]/[OIII] 
diagnostic diagram.   This diagram is chosen because of its high
sensitivity to the ionization level of the gas, whose variation, as proposed above,
can explain  the trends defined by type 2 quasars in those diagnostic diagrams with the highest sensitivity to $U$.  The data for most Seyfert 2s  were collected
from  \cite{kos78}. Some more line ratios were taken from papers on 
individual
sources (e.g. Bennert et al. \citeyear{ben06}, Ferruit et al. \citeyear{fer99} , 
Contini et al. \citeyear{con02}, Allen et al. \citeyear{all99}, Diaz et al.
\citeyear{diaz88}, Kraemer \& Crensaw \citeyear{kra00}). The   radio galaxy data set is the same as in \cite{holt06}. 
None of these line ratios have been corrected for internal reddening.
 Fig. 2 shows that type 2 quasars  overlap in
 the diagram with  radio galaxies with moderate and high ionization level
 (lg($U$)$\ga$-2.6) and with most Seyfert 2 galaxies. This suggests similar  ionization mechanism, ionizing continuum shape, physical properties of
the gas and a similar range of chemical abundances. The lack of type
2 quasars with lg([OIII]/H$\beta)\la$0.5 is due to  the selection
criteria (see Zakamska et al. \citeyear{zak03}).

\begin{figure}
\includegraphics{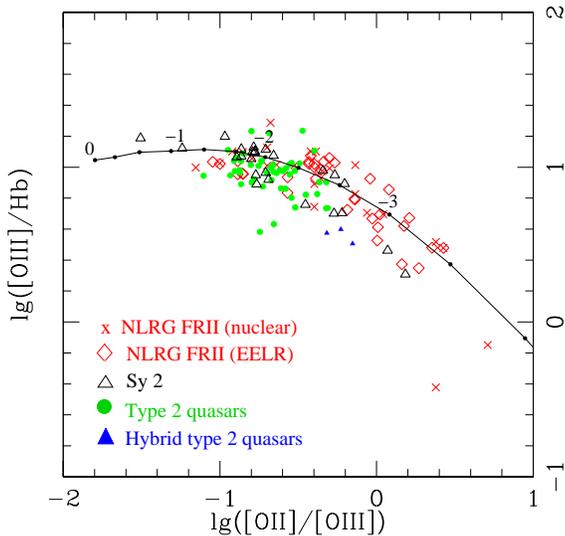}
\vspace{2.8in}
\caption{Comparison among type 2 quasars, Seyfert 2s and narrow line
FRII radio galaxies (nuclear and extended emission line regions, EELR).   
The solid line is the AGN model sequence, as in Fig.~1. The three object classes overlap in the area of the diagram with lg($U$)$\ge$-2.6. The lack of type
2 quasars with lg([OIII]/H$\beta)\la$0.5 is due to the selection
criteria.}
\end{figure}

\item Type 2 quasars are in general clearly separated from HII galaxies (see diagrams $a$, $b$ and $d$). 
This is particularly evident in diagram $d$, involving
the [NII]/H$\alpha$ ratio (see also Fig.~3). This is also evident in Fig.~3, which shows one of the most efficient diagrams at separating active galaxies
and HII galaxies (Baldwin, Philips \& Terlevich \citeyear{bal81}).
The [NII]/H$\alpha$ ratio    was measurable only for 30\% of the objects.
In a few cases, the lines are in a very noisy part
of the spectrum due to strong sky residuals. For most spectra, they are 
outside the observed spectral range. In spite of this, the separation from
HII galaxies (also with scarce [NII]/H$\alpha$ measurements) is evident
and type 2 quasars occupy the area of the diagram where active galaxies
lie (see Fig.~5 in Baldwin, Philips \& Terlevich \citeyear{bal81}).  The  harder continuum in active galaxies,
produces a much deeper partially ionized region in which low ionization lines,
such as [NII] are efficiently excited. Although less evident, a similar
effect accounts for the fact that the sequence defined by HII galaxies
in diagrams $a$ and $b$ is shifted down relative to that defined by the
type 2 quasars.

 \begin{figure}
\includegraphics{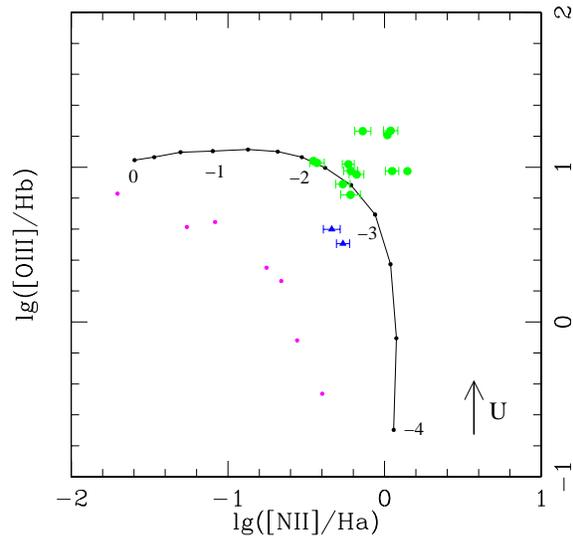}
\vspace{2.8in}
\caption{Type 2 quasars and HII galaxies plotted in Baldwin, Philips \& Terlevich (1982) diagnostic diagram to
classify emission line extragalactic objects. Type 2 quasars are clearly
separated from HII galaxies. They occupy the area of the diagrams where
typical type 2 AGNs lie. Symbols and lines as in Fig.~1.}
\end{figure}

\item  ~[NeV] and/or HeII are detected in all objects, while these lines are 
rarely found in the spectra of star forming galaxies.

\item  The  line luminosities of type 2 quasars are
characteristic of quasars rather than of star forming objects (Zakamska et al. 
\citeyear{zak03}).

\end{itemize}

In spite of the general success of  the standard 
 AGN photoionization models to explain the emission line spectra of type 2 quasars, 
the models encounter some problems, which are exactly those found in studies of other type 2 AGNs (e.g. Groves, Dopita \& Sutherland
\citeyear{grov04}; Binette,
Wilson \& Storchi-Bergmann \citeyear{bin96}; Robinson et al. \citeyear{rob87}).

{\it A. High ionization lines imply higher $U$ values than low ionization lines}

 The diagrams in Fig.~1 show that low and high ionization lines require different
 values of $U$. This is specially clear in diagrams $e$ and $f$ involving the
[NeV] line, where  type 2 quasars are shifted relative to the AGN model sequence.
While the [NeV] ratios imply lg($U$) in the range [-1.8,-1], the implied
range of values in diagrams $a$, $b$ and $c$ is [-2.7,-1.6].
 As for other
AGN types at different redshifts (e.g. Humphrey
et al. \citeyear{hum08}, Binette,
Wilson \& Storchi-Bergmann \citeyear{bin96}), the models that reproduce some of the 
main line ratios under-predict the highest
ionization lines ([NeV]$\lambda$3426 in our case).

Since the faint [NeV] line is detected in most  objects plotted in
Fig.~1, the discrepancy is not the result of different diagrams showing
 different sub-samples.  The problem is found in individual objects.

This cannot be accounted for by reddening effects. Correcting for reddening would
produce  higher  [NeV]/[OII] and [NeV]/[NeIII]  that would result on   larger $U$ values.
On the other hand,  [OII]/[OIII] would also become larger, implying   lower $U$ values. Therefore, the discrepancy would be worse.

{\it B. Too low electronic temperatures}

We have plotted in Fig.~4 the temperature sensitive [OIII]$\lambda\lambda$5007,4959/[OIII]$\lambda$4363 ratio  vs. [OIII]/H$\beta$. The arrows correspond to objects
for which  [OIII]$\lambda$4363 was not detected and only upper limits could
be estimated. For all objects the predicted and measured values of
[OIII]$\lambda\lambda$5007,4959/[OIII]$\lambda$4363  are discrepant by at least
a factor of 2 (a shift of $>$0.3 in lg).  This cannot be accounted for by  measurement errors (see Fig.~4) or reddening effects, since
correcting for this would result on even lower [OIII]$\lambda\lambda$5007,4959/[OIII]$\lambda$4363 ratios.
 
Higher densities can explain low [OIII]$\lambda\lambda$5007,4959/[OIII]$\lambda$4363 values consistent with the data, but produce strong discrepancies with
other line ratios. As an example,  models with  $n$=10$^5$ cm$^{-3}$ produce
lg([OIII]$\lambda\lambda$5007,4959/[OIII]$\lambda$4363)$\le$2, as measured for many 
type 2 quasars. However, for such models
 [OII]/[OIII]$\le$0.02,  [OII]/H$\beta\le$0.3 (much lower than
the measured values, Fig.~1) and [OIII]/H$\beta\ge$18 (higher than measured, Fig.~1).
High densities {\it only}, therefore, in general do not solve the problem.

In the low density limit (Osterbrock \citeyear{ost89}), this discrepancy implies that the
models predict too low  electron temperatures: $T_e\le$11000 K for all models of the $U$ sequence, while the
measured line ratios imply $T_e\ga$15 000 K, being $>$20 000 in several cases, errors considered.
The same problem has been discussed in detail for other type 2 AGNs  (e.g. Binette, Wilson
\& Storchi-Bergmann \citeyear{bin96}, Robinson et al. \citeyear{rob87}).

\begin{figure}
\includegraphics{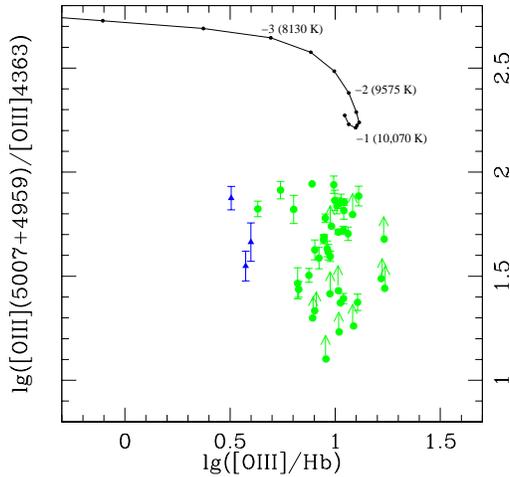}
\vspace{2.8in}
\caption{The [OIII]$\lambda\lambda$5007,4959/[OIII]$\lambda$4363 temperature
sensitive ratio
predicted by the AGN photoionization models is well
above the measured values. Errors 
cannot account for such large discrepancy. Arrows
indicate lower limits. Symbols and lines as in Fig.~1.}
\end{figure}

\vspace{0.2cm}
{\it C. Too small scatter of the HeII/H$\beta$ ratio}
 
The data present a  large HeII/H$\beta$ scatter inconsistent
with the standard AGN sequence (Fig.~1 bottom panel), which is not due to errors in the measurements, neither reddening effects. There are
objects with too high and objects with  too low  HeII/H$\beta$ ratios compared with the model predictions (see  Fig.~5). Such large scatter has been observed also in low $z$ radio galaxies
(e.g. Robinson et al. 1987).

\begin{figure}
\includegraphics{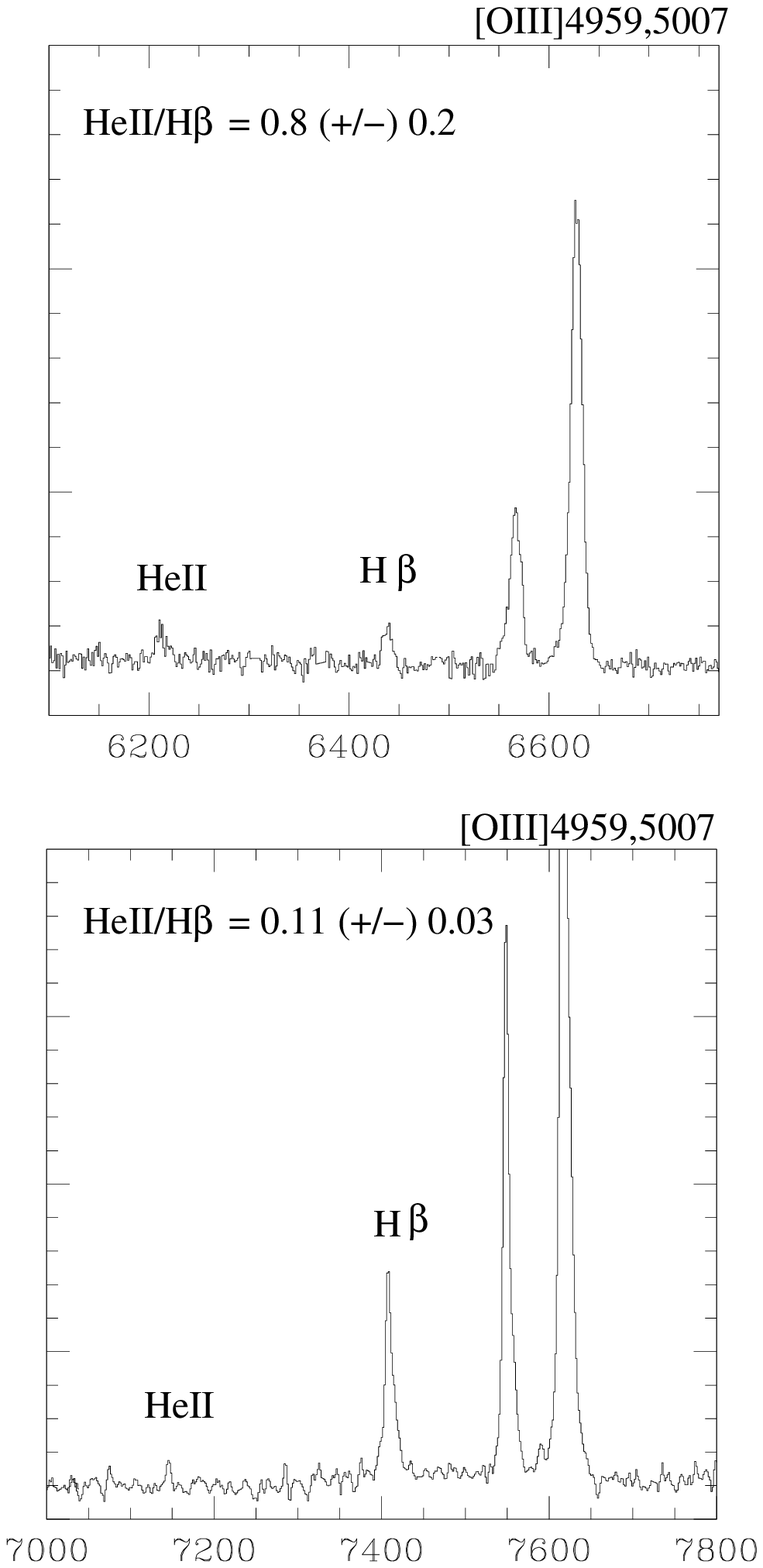}
\vspace{5.55in}
\caption{Examples of objects with very high (top) and very low (bottom) HeII/H$\beta$ ratios ratios.}
\end{figure}

\vspace{0.2cm}
{\it D. Too strong [NII] emission}

The [NII]/H$\alpha$ vs. [OII]/[OIII] diagram (Fig.~1) shows a very large 
scatter in the [NII]/H$\alpha$ ratio
 inconsistent with the standard AGN (solar metallicity) predictions  which again cannot be explained by errors
in the measurements or reddening effects. For a large fraction of objects, the [NII] emission
is too strong.  Given that this ratio is
a direct metallicity indicator, rather than being
a problem for photoionization models, the large range of values suggests that the nitrogen/hydrogen ratio  varies substantially 
within the sample. Nitrogen is likely to be overabundant
in those objects with large [NII]/H$\alpha$ values (e.g. Robinson
et al. \citeyear{rob87}).

\subsection{Alternative AGN scenarios.}

Possible solutions
to these problems have been extensively discussed
in the literature for more than 20 years. We present here a brief summary of the
main alternative AGN photoionization scenarios. 

\begin{itemize}

\item  Mixture of matter and radiation bounded clouds (e.g. Binette, Wilson \& Storchi-Bergmann \citeyear{bin96}; Viegas \& Prieto \citeyear{vie92}).
 Contrary to ionization bounded (IB) clouds, matter bounded (MB)  clouds are not sufficiently thick to absorb all the ionizing radiation.  
In the scenario proposed  by \cite{bin96} the IB component sees a spectrum modified by absorption within the MB component.

\item Locally optimally emitting clouds. The failure of ``simple'' AGN photoionization models to describe the narrow
emission line spectrum of Seyfert 2 galaxies lead \cite{bal95} and 
\cite{ferg97} to propose
an alternative scenario in which the integrated narrow-line spectrum
can be predicted by an integration of an ensemble of clouds with a wide range
of gas densities and distances from the ionizing source with an appropriate covering factor and distribution. For each line, only
a narrow range of density and distance from the continuum source results in
maximum reprocessing efficiency, corresponding to ``locally optimally emitting
clouds''. 
 
\item Dusty, radiation-pressure dominated photoionization models
(Groves, Dopita \& Sutherland \citeyear{grov04}, Dopita et al. \citeyear{dop02}). In these
models, dust and the radiation pressure acting upon it provide the controlling factor in moderating the density, excitation and surface brightness of the photoionized gas. Additionally, photoelectric heating by the dust
modifies the gaseous temperature structure.

\item  \cite{hum08} discuss in detail the need for a mixture of cloud properties (a range in $U$) in
individual objects to explain the UV and optical rest-frame line ratios of high $z$ radio galaxies.

\end{itemize}

%
%

\subsection{Stellar photoionization}

There are 3 objects, highlighted as blue triangles in Fig.~1, which show hybrid properties of both AGNs 
and HII galaxies. Their line ratios in diagrams $a$, $b$ and $c$ 
fit better a classification as HII galaxies.  Their lines are also relatively
narrow compared with most objects in the sample, for which the median value is 530 km s$^{-1}$
while the hybrid objects have FWHM[OIII]$<$470 km s$^{-1}$. They show among the
lowest [OII]/[OIII] and [OIII]/H$\beta$ values in the sample.

On the other hand,  they show some features characteristic of active galaxies: they emit [NeV] and HeII  (Fig.~1) 
and have large line  luminosities 
(L[OIII]$\ga$10$^{42}$
erg s$^{-1}$).  The two objects for which [NII] was measured occupy an intermediate region in the [OIII]/H$\beta$ vs. [NII]/H$\alpha$ diagram (Fig.~3)
 between  HII galaxies and AGNs. We will call
these  hybrid objects.  

Based on the arguments exposed in the previous sections, AGN
photoionization must play an important role in the ionization of the gas in
 type 2 quasars in this sample.  The discrepancies of the standard AGN sequence show
that  a range of ionization and probably physical properties must exist within the type 2 quasar sample.  An internal range of cloud properties (e.g. density range, matter and bounded clouds, ionization level etc)  is also likely to exist in individual objects.
This is only natural, one cannot expect identical
gas and continuum properties in all quasars, or ensembles of identical clouds in individual objects.
As we discussed above,  more sofisticated  models that take this into account solve the problems of the standard AGN sequence.

However, the emission line spectra of 
 the three
hybrid objects suggests that stellar photoionization might also be present
with different degrees of  importance relative to AGN photoionization from object to object.
 Interestingly, although uncommonly high [OIII] luminosities for HII galaxies,
the hybrid  objects are at the lowest end of L[OIII]
values  within the type 2 quasar sample ($\la$2$\times$10$^{42}$ erg s$^{-1}$, while the median value is $\sim$6$\times$10$^{42}$ erg s$^{-1}$). This tentatively suggests
that low line luminosities might be associated with relatively stronger stellar photoionization. 

\begin{figure*}
\includegraphics{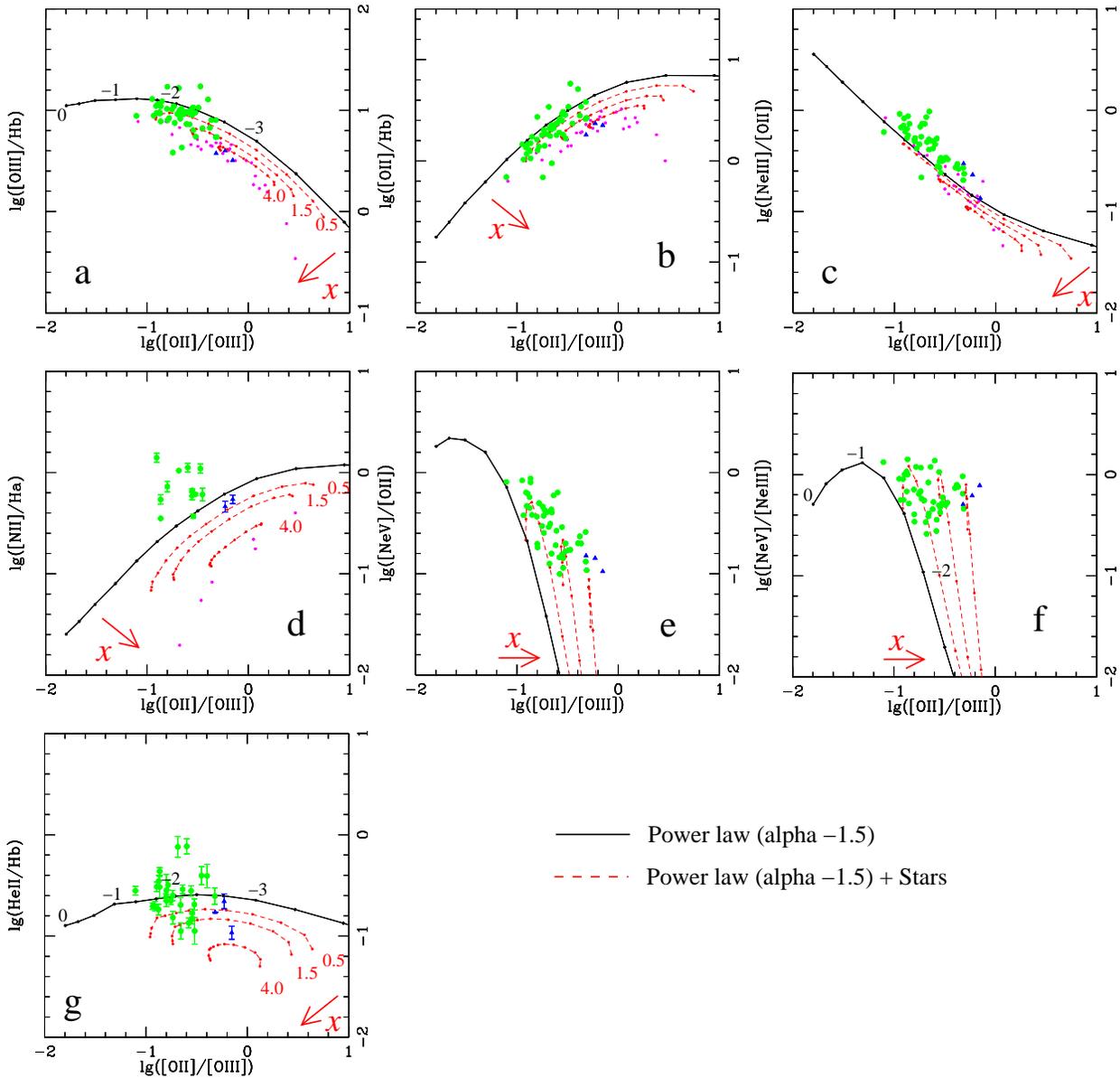}
\vspace{6.5in}
\caption{As Fig.~1, adding the hybrid AGN+stellar models (red, dashed lines) for different
$x$ values shown in red (0.5, 1.0 and 4.0). 
The AGN sequence (black solid line) corresponds to $x$=0, i.e., the line fluxes are emitted by gas purely photoionized
by the AGN. The arrows indicate the sense of increasing $x$.}
\end{figure*}

We investigate next whether a contribution of stellar photoionized gas
could solve the problems of the standard AGN  models discussed above. 
For simplicity, 
we have assumed that the total flux of a given
line is due to the added contribution of the flux emitted by an AGN photoionized component (represented by the standard AGN sequence) 
plus the flux emitted by a stellar photoionized component. To represent the spectrum
emitted by this gas, we have used the real spectra
of a variety of HII galaxies in  \cite{ter91} catalogue.  We find from
this study that the stellar ionized component must have
[OII]/[OIII]$\ga$1 in order to shift the models in the right direction in
the diagrams.

As an example we take UM448. The reason to choose this particular 
HII galaxy is  that it fulfills the  [OII]/[OIII]$\ga$1
requirement and has reported measurements of
 [OIII]$\lambda$4363.  Unfortunately, as for all other objects in the catalogue
with [OII]/[OIII]$\ga$1,  HeII is not measured.

The flux of any line $Flux^{tot}$ relative to H$\beta^{tot}$ is given by: 

$$\frac{Flux^{tot}}{H\beta^{tot}} = \frac{Flux^*+Flux^{AGN}}{H\beta^*+H\beta^{AGN}}$$

where   $Flux^*$ and H$\beta^*$ are respectively the flux
of the line and the flux of H$\beta$ emitted by the stellar ionized gas.  
$Flux^{AGN}$ and H$\beta^{AGN}$ are the flux of the line and
that of H$\beta$ emitted by gas 
ionized by the AGN.

Let us define $x=\frac{H\beta^*}{H\beta^{AGN}}$ and rearrange:

$$\frac{Flux^{tot}}{H\beta^{tot}} =\frac{\frac{Flux^{AGN}}{H\beta^{AGN}} + x ~ \frac{Flux^*}{H\beta^*}}{1~+ ~ x} $$

We assume [NeV]$^*$/H$\beta^{*}<<$ [NeV]$^{AGN}$/H$\beta^{AGN}$.  When detected,  HeII$^*$/H$\beta^*$ in HII galaxies is often 
in the range $\sim$0.02-0.1 (values as high as 0.4 are also possible, but
are more typical of objects with very small [OII]/[OIII]$\la$0.2 values). 
We will assume HeII$^*$/H$\beta^*$=0.04. Using a different value within the expected range does not 
change our conclusions. 

We now create hybrid models by adding the stellar emission line spectrum to
the AGN standard sequence of models and changing the $x$ value from sequence
to sequence (see equation above). The ionization parameter $U$ of the AGN photoionized gas (as in
the standard AGN sequence) changes along each sequence. The results are shown in Figs.~6, 7 and 8 as red dashed-lines.

We find that by varying the relative contribution of the stellar to the AGN
photoionized gas:

\begin{itemize}

\item The discrepancy affecting the high and low ionization lines disappears (Fig.~6, panels $e$ and $f$).

\item The large scatter observed in the HeII/H$\beta$ diagram towards low values is now reproduced (Fig.~6, panel $g$).

\item  The highest HeII/H$\beta$ values ($\sim$0.8 for two objects) cannot
be reproduced by the models   (Fig. 6, panel $g$).To solve this problem with
the stellar photoionized component, this should show similarly high HeII/H$\beta$ ratios, which are extreme
even for active galaxies. We 
notice this is also a problem for
 Groves, Dopita \& Sutherland (2004)
 models, which require extremely high gas metallicities $\sim$4 $Z_{\odot}$. 
No models in Ferguson et al. (\citeyear{ferg97}) reach these high HeII/H$\beta$ values
either (see Fig.~3b in their paper). A  strong contribution
of  matter bounded clouds to the integrated spectrum solves the problem (Binette et al. \citeyear{bin96}).
 Given the scarcity of these extreme  HeII/H$\beta$ ratios in type 2 AGNs in general,
we wonder whether these two specific type 2 quasars are particularly rare. 

\item Although the temperature problem is partially alleviated (Fig.~7), the hybrid models cannot
explain the small [OIII]$\lambda\lambda$5007,4959/[OIII]$\lambda$4363 measured in the quasar sample.
Given
the saturation of this line ratio
 at high electron temperatures (Osterbrock 1989), not even unrealistically high
electron temperatures of a dominant stellar photoionized gas would solve the problem.
Notice that  Binette et al. (1997) warned about the need for higher densities  if the NLR  dominates the emission, which would result
in lower [OIII]$\lambda\lambda$5007,4959/[OIII]$\lambda$4363 ratios for a given electron temperature. 

\item The [NII]/H$\alpha$ ratio is still too low compared with the observations. Adding the stellar
ionized component only makes things worse (Fig.~6, panel $d$ and Fig.~8). However, as explained
before, this could  be suggestive of an overabundance of nitrogen.
\end{itemize}

Another test for the stellar photoionization scenario consists of checking whether
the measured continuum level is consistent with that
expected for the stellar population responsible for  H$\beta^*$.
 Let us consider the  objects
with the largest $x$ values. This is the case of the hybrid objects, for which according to the exercise presented here,
$x\sim$4 (see Fig.~6). For these three objects
the H$\beta$ luminosity is in the range $\sim$3-6$\times$10$^{41}$ erg s$^{-1}$. We will assume
L(H$\beta$)=4.5$\times$10$^{41}$ erg s$^{-1}$. If 80\% is due to stellar photoionized gas
then L(H$\beta^*)$=3.6$\times$10$^{41}$ erg s$^{-1}$, rather similar to that of UMM448,
 L(H$\beta$)=7.9$\times$10$^{40}$ erg s$^{-1}$. 
The rest frame H$\beta$  EW in the hybrid type 2 quasars is in the range $\sim$32-38 \AA, implying EW values
for H$\beta^*\sim$25-30 \AA, 
quite similar to that measured in UM448 (43 \AA, Terlevich et al. \citeyear{ter91}).
 Therefore, for the hybrid objects the H$\beta^*$  luminosities and  equivalent widths expected from stellar photoionization are consistent
with those measured for  HII galaxies.

We must not forget that the continuum emitted by type 2 quasars is not necessarily stellar so that the measured level is an upper limit
to the stellar continuum level and the EW values discussed above for H$\beta^*$ relative to the stellar continuum are lower limits.
A more strict analysis would require a quantification of other possible contaminants, such as   scattered
light from the hidden AGN. Polarimetric information would be needed to characterize the nature of the continuum. 
 Lacking this information, we can only say that the continuum level detected from the hybrid objects,
where the stellar contribution is probably highest, is consistent with that expected from H$\beta^*$.

While this is a  simplistic exercise and one cannot expect
all type 2 quasars to have identical spectra of the stellar ionized gas,
it suggests nevertheless that adding a varying contribution of stellar photoionized gas works in the right direction 
to solve most of the problems affecting the standard AGN sequence. The temperature problem remains
and a more sofisticated scenario with a  range of gas densities or the presence of a matter bounded
component might be a viable solution.

Some studies suggest that the [OIII]$\lambda$5007 emission line is an unbiased indicator of the
intrinsic optical-UV luminositiy of both type 1 quasars and radio galaxies.  (Simpson 1998). According to the results above, 
it is possible that [OIII] has a strong stellar contribution in 
   a fraction of type 2 quasars.
Let us estimate the fraction of [OIII] flux originated by stars in the hybrid objects. We know that:

$$\frac{[OIII]^{tot}}{H\beta^{tot}} =\frac{[OIII]^{tot}}{[OIII]^*}\frac{[OIII]^*}{H\beta^*}\frac{H\beta^*}{H\beta^{tot}}$$ 

using  $\frac{[OIII]^{tot}}{H\beta^{tot}}$ in the range 3.2-3.9 as measured for the hybrid objects,
$\frac{OIII^*}{H\beta^*}$=2.9 (as for UM448) and $\frac{H\beta^*}{H\beta^{tot}}$=0.8 (since $x$=4) and rearranging,
$\frac{[OIII]^*}{[OIII]^{tot}}\sim$59-72\%. This suggests that there could be a fraction of type 2 quasars in which [OIII] is not a reliable indicator of AGN
power. Other type of studies should be performed to investigate this issue  more carefully (e.g. Simpson \citeyear{sim98}).

\begin{figure}
\includegraphics{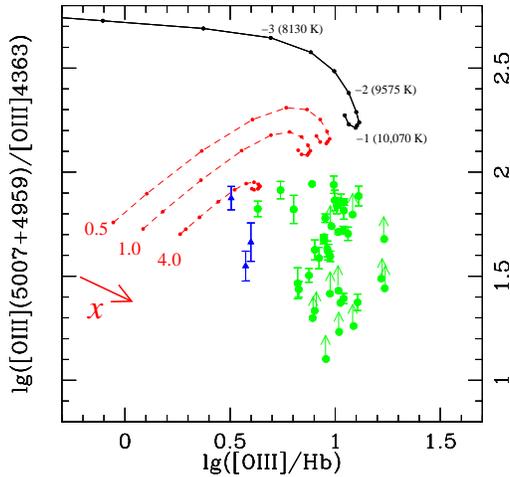}
\vspace{2.8in}
\caption{As Fig.~4, adding the hybrid AGN+stellar models (red, dashed lines) for different
$x$ values shown in red (0.5, 1.0 and 4.0).}
\end{figure}

 \begin{figure}
\includegraphics{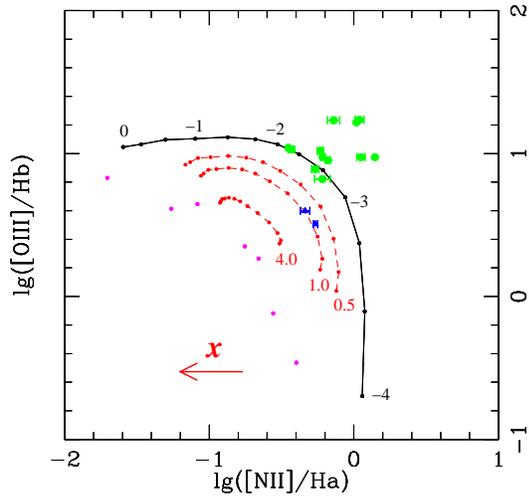}
\vspace{2.8in}
\caption{As Fig.~3 with the hybrid models.}
\end{figure}

\section{Discussion and conclusions}

We have compared the optical emission lines ratios of type 2 quasars from
\cite{zak03}
with standard AGN photoionization model predictions, Seyfert 2s, HII galaxies,
and narrow line FRII radio galaxies.  Moderate to high ionization narrow line radio galaxies and Seyfert 2s are indistinguishable from type 2 quasars based 
on their optical line ratios. The standard AGN photoionization models, valid for other type 2 AGNs, can reproduce successfully the loci and trends
of type 2 quasars in some of the  main diagnostic diagrams.
On the other hand, these models are not exempt of problems and the discrepancies with the data
are the same encountered for other type 2 AGNs.  The comparison between models and data  suggests
that  a range of ionization and probably physical properties must exist within the type 2 quasar sample.  An internal range of cloud properties (e.g. varying density) must also exist in individual objects.
This is only natural, as one cannot expect identical
gas and continuum properties (e.g. ionizing luminosity) in all quasars, or ensembles of identical clouds in individual objects.

Realistic models must take this into account. Possible solutions which have been extensively discussed in 
 the literature are locally optimally emitting clouds, a mixture of matter
and ionization bounded clouds, dusty, radiation-pressure dominated models or a mixture of clouds with different
$U$ values.

 The relevant role played by AGN photoionization is not surprising, 
since \cite{zak03} selected objects with properties characteristic of active galaxies. However, based on the lack of correlation between [OIII] and radio luminosities, other authors have suggested that an important fraction
of type 2 quasars might not be dominated by AGN activity but by star
formation (Vir Lal \& Ho \citeyear{vir07}; see also Kim et al. \citeyear{kim06}).

We have found that stellar photoionization is obvious  in a small fraction of objects (3 out of 50)
which are characterized by low [OIII] luminosities and large [OII]/[OIII] ratios compared with most type 2 quasars in the sample.

 \cite{zak03} sample is strongly biased towards objects with  high line luminosities ( L[OIII]$>$10$^{42}$  erg s$^{-1}$).
 L[OIII] 
can be as low as $\sim$10$^{40}$ erg s$^{-1}$  in radio-quiet type 1 quasars and
narrow line radio galaxies (e.g. Bennert et al. \citeyear{ben02}, Tadhunter et al. \citeyear{tadh98}). There must be many type 2 quasars
with L[OIII] in the range $\sim$10$^{40-42}$ erg s$^{-1}$. The hybrid objects discussed here,
where stellar photoionization is obvious,
have among the lowest L[OIII] values within the type 2 quasar sample of \cite{zak03}. This tentatively suggests that stellar photoionization could be  relatively more important
in type 2 quasars  with lower [OIII] luminosities. If these objects have
lower power AGNs (e.g. Simpson \citeyear{sim98}), it would seem plausible
 to find a relatively higher
 contribution of the stellar photoionized gas to the observed emission line spectrum. The fraction of hybrid objects, therefore, could be  much larger, if  
lower L[OIII]  values were considered.

 Since star formation has been found in differents classes of  type 2 AGNs  at different $z$ (e.g. Holt et al. \citeyear{holt07}, Tadhunter et al. \citeyear{tadh05}, Alonso-Herrero et al. \citeyear{al08}) and  in at least a fraction
of type 2 quasars  (Mart\'\i nez-Sansigre et al. \citeyear{mar08}, Lacy et al. \citeyear{lacy07}), stellar+AGN photoionization is a plausible scenario. Inspired by these arguments, we have  explored an alternative scenario to pure AGN photoionization in which a varying contribution of stellar ionized gas is added to
the line fluxes in type 2 quasars. Although the hybrid models presented here are rather simplistic and no reliable  quantitative results
can be extracted about the relative importance of stellar vs. AGN photoionization, they reproduce the
type 2 quasar ratios (and type 2 AGN in general) quite successfully (better than the standard AGN sequence). This suggests that
stellar photoionization might also be present in many type 2 quasars, in addition to AGN photoionization. 
 Given that other type 2 AGNs
have similar line ratios, this applies as well in those cases. On the other hand, and contrary to the more sofisticated AGN models discussed in the literature,
the hybrid models cannot solve the ``temperature problem'' (see \S3.1 $B$). 
Regarding all other line ratios, given the strong degeneracy with more sofisticated AGN photoionization models, it is not possible
to favour one scenario or another in terms of the line ratios.

Other sources of information would be very valuable to test whether
stellar photoionization is present in  type 2 quasars and characterize  its importance relative to the AGN photoionization
(e.g. evidence for extended star formation,  possible correlations between  star formation indicators and the line ratios 
which require a stronger contribution of stellar ionized gas, lower continuum polarization level in objects
with hints of stellar photoionization from the line ratios, etc). If the emission lines, in particular [OIII]$\lambda$5007 
have a strong contribution of stellar
photoionized gas, the [OIII] line might not be a good indicator of AGN power.

\section*{Acknowledgments}
The authors thank to an anonymous referee  for useful comments which helped to improve the paper. The work by MVM has been funded with support from the Spanish Ministerio de 
Educaci\'on y Ciencia through the grants AYA2004-02703 and 
AYA2007-64712, and co-financed with FEDER funds.
Thanks to Joanna Holt for providing the data set on radio galaxies.
LB was supported by the CONACyT grant J-50296


\begin{thebibliography}{}

\bibitem[\protect\citeauthoryear{Allen et al.}{1999}]{all99} 
Allen M., Dopita M., Tsvetanov Z., Sutherland R., 1999, ApJ, 511, 686

\bibitem[\protect\citeauthoryear{Alonso-Herrero et al.}{2008}]{al08} 
Alonso-Herrero A. Pérez-Gonz\'alez P. G., Rieke G., 
Alexander D. M., Rigby J., Papovich C., Donley J., Rigopoulou D.,
2008,  ApJ, 277, 127


\bibitem[\protect\citeauthoryear{Anders \& Grevesse}{1989}]{anders89} 
Anders  E. \& Grevesse N., 1989, Geochim. Cosmochim. Acta, 53, 197
 
\bibitem[\protect\citeauthoryear{Antonucci}{1993}]{ant93} 
Antonucci R., 1993, ARA\&A, 31, 473

\bibitem[\protect\citeauthoryear{Baldwin, Philips \& Terlevich}{1982}]{bal81} 
Baldwin J., Philips M. \& Terlevich R., 1981, PASP, 93, 5

\bibitem[\protect\citeauthoryear{Baldwin et al.}{1995}]{bal95} 
Baldwin J., Ferland G., Korista K., Verner D., 1995, ApJ, 455, L119


\bibitem[\protect\citeauthoryear{Bennert et al.}{2002}]{ben02} 
Bennert N., Falcke H., Shchekinov Y., Wilson A., Wills B., 2002,
 ApJ, 574, L105

\bibitem[\protect\citeauthoryear{Bennert et al.}{2006}]{ben06} 
Bennert N., Jungwiert  B., Komossa  S., Haas  M., Chini  R., 2006,
A\&A, 2006, 456, 953
 
\bibitem[\protect\citeauthoryear{Binette, Dopita \& Tuohy}{1985}]{bin85} 
Binette L., Dopita M. \& Tuohy I.R.; 1985, ApJ, 297, 476

\bibitem[\protect\citeauthoryear{Binette, Wilson \& Storchi-Bergmann}{1996}]{bin96} 
Binette L., Wilson A., Storchi-Bergmann T., 1996, A\&A, 312, 365

\bibitem[\protect\citeauthoryear{Clark et al.}{1998}]{clark98} 
Clark N. E., Axon D. J., Tadhunter C. N., Robinson A., O'Brien P.,
1998, MNRAS, 494, 546

\bibitem[\protect\citeauthoryear{Contini et al.}{2002}]{con02} 
Contini M., Radovich M., Rafanelli P., Richter G.M., 2002, ApJ, 572, 124

\bibitem[\protect\citeauthoryear{D\'\i az et al.}{1988}]{diaz88} 
Diaz A.I., Prieto M.A., Wamsteker W., 1988, A\&A, 195, 53

\bibitem[\protect\citeauthoryear{Dopita \& Sutherland}{1996}]{dop96} 
Dopita M., Sutherland R., 1996, ApJS, 102, 161

\bibitem[\protect\citeauthoryear{Dopita et al}{2002}]{dop02} 
Dopita M., Groves B., Sutherland R., Binette L., Cecil G., 2002,
ApJ, 572, 753

\bibitem[\protect\citeauthoryear{Ferguson et al.}{1997}]{ferg97} 
Ferguson D. H., Korista K., Baldwin J., Ferland G., 1997, ApJ, 487

\bibitem[\protect\citeauthoryear{Ferruit et al.}{1997}]{fer97} 
Ferruit P., Binette L., Sutherland R. S., Pecontal E., 1997,
A\&A, 322, 73

\bibitem[\protect\citeauthoryear{Ferruit et al.}{1999}]{fer99} 
Ferruit P., Wilson A., Whittle M., Simpson C., Mulchaey J., Ferland G., ApJ,
1999, 523, 147

\bibitem[\protect\citeauthoryear{Groves, Dopita \& Sutherland}{2004}]{grov04} 
Groves B., Dopita M., Sutherland R., 2004, ApJSS, 153, 75

\bibitem[\protect\citeauthoryear{Holt}{2006}]{holt06} 
Holt J., 2006, PhD thesis, Univ. of Sheffield

\bibitem[\protect\citeauthoryear{Holt et al.}{2007}]{holt07} 
Holt J., Tadhunter C.N., Gonz\'alez Delgado R., Inskip K., Rodr\'\i guez J., Emmonts B.,
Morganti R., Wills K., 2007, MNRAS, 381, 611


\bibitem[\protect\citeauthoryear{Humphrey et al.}{2008}]{hum08} 
Humphrey A.,  Villar-Mart\i n M., Vernet J., Fosbury R., di Serego Alighieri S., Binette L., 2008, MNRAS, 383, 11

\bibitem[\protect\citeauthoryear{Kim, Ho \& Im}{2006}]{kim06} 
Kim M., Ho L.C., Im M., 2006, ApJ, 642, 702

\bibitem[\protect\citeauthoryear{Koski}{1978}]{kos78} 
Koski A., 1978, ApJ, 223, 56

\bibitem[\protect\citeauthoryear{Kraemer \& Crensaw}{2000}]{kra00} 
 Kraemer S., Crensaw D., 2000, ApJ, 544, 763

\bibitem[\protect\citeauthoryear{Lacy et al.}{2007}]{lacy07} 
Lacy M., Sajina A., Petric A., Seymour N., Canalizo G., Ridgway S., Armus L., Storrie-Lombardi L.,
2007, ApJ, 669L, 61

\bibitem[\protect\citeauthoryear{Mart\'\i nez-Sansigre et al.}{2005}]{mar05} 
Mart\'\i nez-Sansigre A., Rawlings S., Lacy M., Fadda D., Marleau F., Simpson
C., Willott C., Jarvis M., 2005, Nature, 436, 666

\bibitem[\protect\citeauthoryear{Mart\'\i nez-Sansigre et al.}{2008}]{mar08} 
Mart\'\i nez-Sansigre A., Lacy M., Sajina A., Rawlings S., 2008,  ApJ, 
674, 676 

\bibitem[\protect\citeauthoryear{Osterbrock}{1989}]{ost89} 
Osterbrock D.E., 1989,  Astrophysics of Gaseous Nebulae and Active Galac-
    tic Nuclei. University Science Books, Mill Valley, CA


\bibitem[\protect\citeauthoryear{Ptak et al.}{2006}]{ptak06} 
Ptak A., Zakamska N., Strauss M., Krolik J., Heckman T., Schneider D.,
Birnkmann J., 2006, ApJ, 637, 147 

\bibitem[\protect\citeauthoryear{Reyes et al.}{2008}]{rey08} 
Reyes R., Zakamska N., Strauss M., Green J., Krolik J., Richards
G., Anderson s., Schneider D., 2008, submitted to AJ (arXiv:0801.1115)

\bibitem[\protect\citeauthoryear{Robinson et al.}{1987}]{rob87} 
Robinson A., Binette L., Fosbury R.A.E., Tadhunter C.N., 1987, MNRAS, 227. 97


\bibitem[\protect\citeauthoryear{Simpson}{1998}]{sim98} 
Simpson C., 1998, MNRAS, 297, L39

\bibitem[\protect\citeauthoryear{Szokoly et al.}{2004}]{szo04} 
Szokoly G. P., Bergeron J., Hasinger G. et al., 2004, ApJS, 2004, 155, 271

\bibitem[\protect\citeauthoryear{Tadhunter, Fosbury \& Quinn}{1989}]{tadh89} 
Tadhunter C.N., Fosbury R.A.E., Quinn P.J., 1989, MNRAS, 240, 225



\bibitem[\protect\citeauthoryear{Tadhunter et al.}{1998}]{tadh98} 
Tadhunter C.N., Morganti R., Robinson A., Dickson R., Villar-Mart\'\i n M., 
Fosbury R.A.E., 1998, MNRAS, 298, 1035

\bibitem[\protect\citeauthoryear{Tadhunter}{2002}]{tadh02} 
Tadhunter C.N., 2002, Revista Mexicana de Astronom\'\i a y Astrof\'\i sica (Serie de Conferencias), 13, 213-221

\bibitem[\protect\citeauthoryear{Tadhunter et al.}{2005}]{tadh05} 
Tadhunter C., Robinson T., González Delgado R., Wills K., Morganti R., 2005, MNRAS, 356, 480

\bibitem[\protect\citeauthoryear{Terlevich et al.}{1991}]{ter91} 
Terlevich R., Melnick J., Masegosa J., Moles M., Copetti M., 1991, A\&AS,
91, 285

\bibitem[\protect\citeauthoryear{Viegas  \& Prieto}{1992}]{vie92} 
Viegas S., Prieto A., 1992, MNRAS, 258, 483

\bibitem[\protect\citeauthoryear{Villar-Mart\'\i n et al.}{1999}]{vm99} 
Villar-Martín M., Tadhunter C., Morganti R., Axon D., Koekemoer A., 1999,
MNRAS, 307, 24

\bibitem[\protect\citeauthoryear{Villar-Mart\'\i n et al.}{2005}]{vm05} 
Villar-Mart\'\i n M., Tadhunter C., Morganti R., Holt J., 2005, MNRAS,
 359, 5

\bibitem[\protect\citeauthoryear{Vir Lal \& Ho}{2007}]{vir07} 
Vir Lal D.,  Ho L.C., 2007, in press. To appear in "The Central Engine of Active Galactic Nuclei", ed. L. C. Ho and J.-M. Wang (San Francisco: ASP); 
arXiv:0706.0148

\bibitem[\protect\citeauthoryear{York et al.}{2000}]{york00} 
York D., Adelman J., Anderson J. et al., 2000, AJ, 120, 1579

\bibitem[\protect\citeauthoryear{Zakamska et al.}{2003}]{zak03} 
Zakamska N., Strauss M., Krolik J. et al. 2003, AJ, 126, 2125

\bibitem[\protect\citeauthoryear{Zakamska et al.}{2004}]{zak04} 
Zakamska N., Strauss M., Heckman T., Ivezi\'c Z., Krolik J.,  2004, 
AJ, 128, 1002

\bibitem[\protect\citeauthoryear{Zakamska et al.}{2006}]{zak06} 
Zakamska N., Strauss M.,  Krolik J. et al. 2006, 
AJ, 132, 1496

\bibitem[\protect\citeauthoryear{Zhang, Dultzin-Hacyan \& Wang}{2007}]{zh07} 
Zhang Z., Dultzin-Hacyan D. \& Wang T., 2007, MNRAS, 377, 1215 

\end{thebibliography}
\end{document}